\documentclass{article}
\usepackage{spconf,amsmath,graphicx}
\usepackage{multirow,soul,xcolor,boldline,hhline,hyperref, enumitem,wrapfig, cite}
\usepackage{pifont}% http://ctan.org/pkg/pifont
\usepackage{graphicx,multirow,soul,boldline,hyperref,subcaption,xcolor,soul,flushend}
\usepackage{booktabs} % for borders and merged ranges

\usepackage{changepage,threeparttable} % for wide tables
\usepackage{makecell}
\usepackage{xspace}

\newif\ifarxiv
\arxivfalse

\newif\ifdraft
\draftfalse

\definecolor{mygray}{gray}{0.7}
\definecolor{dkgreen}{RGB}{0,179,36}
\definecolor{dkorange}{RGB}{230,115,0}

\newcommand{\baselineA}{fine-tuning baseline\xspace}
\newcommand{\baselineB}{context injection baseline\xspace}
\newcommand{\method}{context-aware fine-tuning\xspace}

\ifdraft

\newcommand{\suwondel}[1]{\textcolor{green}{{\st{#1}}}}
\newcommand{\kledit}[1]{\textcolor{blue}{{#1}}}
\newcommand{\klremove}[1]{\textcolor{blue}{{\st{#1}}}}
\else

\newcommand{\suwondel}[1]{{}}
\newcommand{\kledit}[1]{\textcolor{black}{{#1}}}
\newcommand{\klremove}[1]{{}}
\fi

\title{Context-aware fine-tuning of self-supervised speech models}

\name{Suwon Shon$^1$, Felix Wu$^1$, Kwangyoun Kim$^1$, Prashant Sridhar$^1$, Karen Livescu$^{2}$, Shinji Watanabe$^{3}$}
\address{$^1$ASAPP \ \ \ \ \   $^2$Toyota Technological Institute at Chicago \ \ \ \ $^3$Carnegie Mellon University}

\begin{document}
\ninept
\maketitle

\begin{abstract}
% \vspace{-1mm}
Self-supervised pre-trained transformers 
have improved the state of the art on a variety of speech tasks.
Due to the quadratic time and space complexity of self-attention, they usually operate at the level of relatively short (e.g., utterance) segments.
In this paper, we study the use of context, \textit{i.e.}, surrounding segments, during fine-tuning and propose a new approach called \method. 
We attach a context module on top of the last layer of a pre-trained model to encode the whole segment into a context embedding vector which is then used as an additional feature for the final prediction. 
During the fine-tuning stage, we introduce an auxiliary loss that encourages this context embedding vector to be similar to context vectors of surrounding segments.
This allows the model to make predictions without access to these surrounding segments at inference time and requires only a tiny overhead compared to standard fine-tuned models.
We evaluate the proposed approach using the SLUE and Libri-light benchmarks for several downstream tasks: Automatic speech recognition (ASR), named entity recognition (NER), and sentiment analysis (SA).  The results show that \method not only outperforms a standard \baselineA but also rivals a strong \baselineB that uses neighboring speech segments during inference.

\end{abstract}


\section{Introduction}
\label{sec:intro}
Self-supervised pre-training has significantly improved the state of the art in speech processing~\cite{mohamed2022self,baevski2020wav2vec,hsu2021hubert_TASLP,oord2018representation,chung2019unsupervised,liu2020mockingjay}, and fine-tuning pre-trained models has become the \textit{de facto} way to achieve good results on downstream speech recognition and understanding tasks\kledit{~\cite{kahn2020libri,wang2021voxpopuli,shon2022slue,yang2021superb,le2022lebenchmark,pasad2021use_NAACL}}. 
% These large pre-trained models usually uses Transformer architecture which encodes context information based on self-attention mechanism.
Despite their success, because of the quadratic complexity in their self-attention layers, these pre-trained and fine-tuned models usually operate at the utterance level, which limits their ability to make decisions with context information.

% To utilize the context information, many efforts have been made in both natural language processing (NLP) and speech fields.
% In NLP, the context goes beyond utterances because the text are more compact and is usually used for pre-training objectives.
Utilizing nearby utterances in the training objective was first proposed and studied in natural language processing (NLP), starting with the skip-thought vectors of Kiros et al.~\cite{kiros2015skip} and continuing with more recent models like BERT~\cite{devlin2018bert} and ALBERT\kledit{~\cite{lan2019albert_ICLR}}.  All of these models are trained with a loss that considers the past/future text content, either by requiring the model to generate the context, to discriminate between correct/incorrect next sentences, to reorder sentences relative to each other, or to embed previous text segments.
%Kiros et al.~\cite{kiros2015skip} introduced skip-thought vectors which is an LSTM encoder that produces a hidden vector that can be used as the initial state of another two LSTMs to generate the previous and the next utterances, respectively. More recently, BERT~\cite{devlin2018bert} is pre-trained on a next sentence prediction (NSP) objective where the model has to discriminate whether the second input sequence is actually the next sentence in the original document or a random sentence by reasoning about the relationships between the sentences.ALBERT~\cite{lan2019albert} introduces a more difficult variant, sequence ordering task, where the model has to predict whether two sentences come in the correct order or they are swapped.  Transfomer-XL~\cite{dai2019transformer} introduces a different way rather than reasoning via objective. It tries to re-use the hidden state from the previous segment to learn longer contextual information from text articles.

In the speech processing field, similar studies have been done for Spoken Language Understanding (SLU) tasks~\cite{sunder2022towards,ganhotra2021integrating_IS}, using models that consume the text from the entire conversation history. For ASR, many works focus on learning frame-level short-term contextual information\kledit{~\cite{oord2018representation,tsunoo2019transformer,chung2019unsupervised,baevski2020wav2vec,kim2021multi,an2022cuside_IS}}. For longer-term contextual information, several studies~\cite{hori2021advanced,tomashenko2020dialogue,wei2022conversational,wei2022leveraging,kim2019acoustic,kim2019cross} utilize the previous speech segment or text for current speech segment decoding. Regardless of the performance improvement, one limitation of these approaches is that the computation cost is high, since previous segments need to be encoded. In addition, the context is restricted to previous speech segments. How to better utilize the surrounding segments remains an important question in speech processing.

% In speech, utilizing the acoustic inputs across multiple speech segments is rarely studied~\cite{hori2021advanced}. More works focused on learning frame-level short-term contextual information~\cite{oord2018representation,tsunoo2019transformer,chung2019unsupervised,baevski2020wav2vec,kim2021multi,an2022cuside}; others rely on text-form inputs from previous speech segments~\cite{kim2019acoustic,kim2019cross}. To the best of our knowledge, the only work that falls into this category is context-expanded transformers\cite{hori2021advanced} which encode multiple previous segments together while transcribing the current segment.
% How to better utilize the surrounding segments remains an important question in speech processing.

In this paper, we propose \method, which utilizes the previous or future speech segments as contextual information during only the fine-tuning stage but not inference.
Specifically, we introduce a context generator module that encodes the previous/future speech segments into a context embedding vector or generates a context embedding from the current speech segment. The context embedding is then concatenated with the speech representation at each frame to make predictions.
During fine-tuning, a context loss is added to ensure that the context embedding vector of the current speech segment is similar to the context embedding vector of neighboring speech segments.
This \method encourages the model to learn to predict contextual information from nearby speech segments. There is no limitation that the nearby speech segments be previous or future segments, since they are needed only for fine-tuning.
We conduct experiments on the SLUE~\cite{shon2022slue} and Libri-light~\cite{kahn2020libri} benchmarks to evaluate ASR, named entity recognition (NER) and sentiment analysis (SA).
Our experiments show that \method matches the high accuracy of a much slower \baselineB (using multiple speech segments) while enjoying the same fast inference as a standard \baselineA.

\begin{figure*}[th]
     \centering
    \begin{subfigure}[b]{0.2\textwidth}
         \centering
         \includegraphics[width=\textwidth]{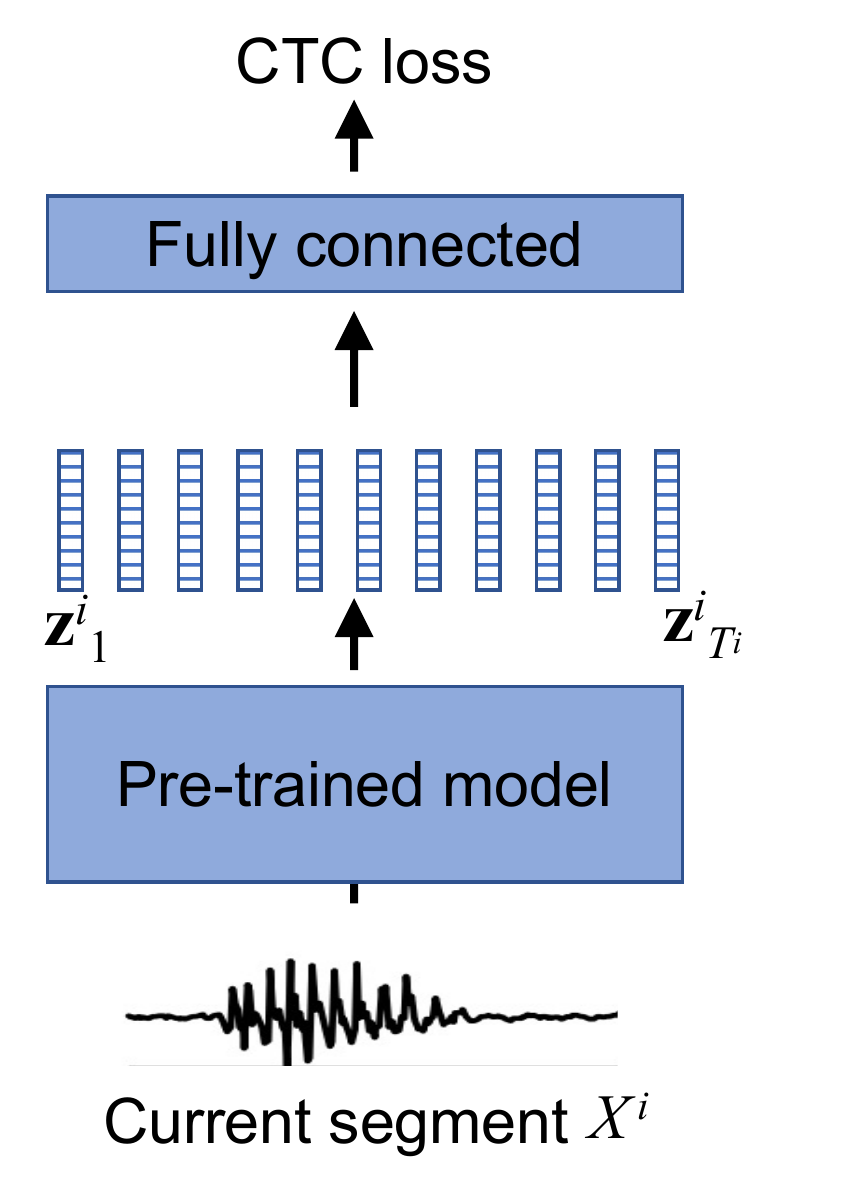}
         % \vspace{-0.5cm}
         \caption{Fine-tuning baseline}
         \label{subfig:concept_baselineA}
     \end{subfigure}
     \hspace{-5pt}
    \begin{subfigure}[b]{0.37\textwidth}
         \centering
         \includegraphics[width=\textwidth]{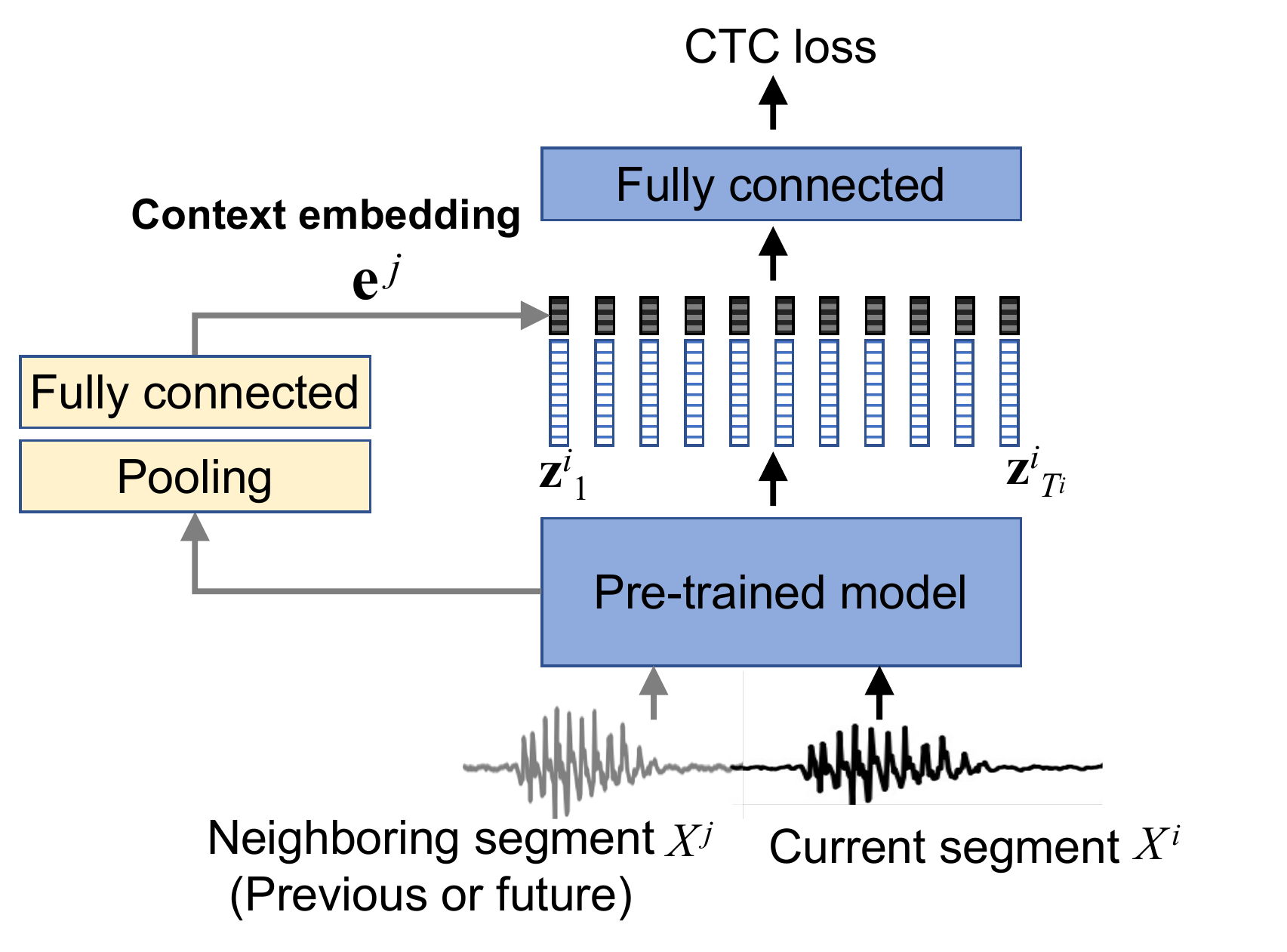}
         % \vspace{-0.5cm}
         \caption{Context injection baseline}
         \label{subfig:concept_baselineB}
     \end{subfigure}
     \hspace{-5pt}
     \begin{subfigure}[b]{0.39\textwidth}
         \centering
         \includegraphics[width=\textwidth]{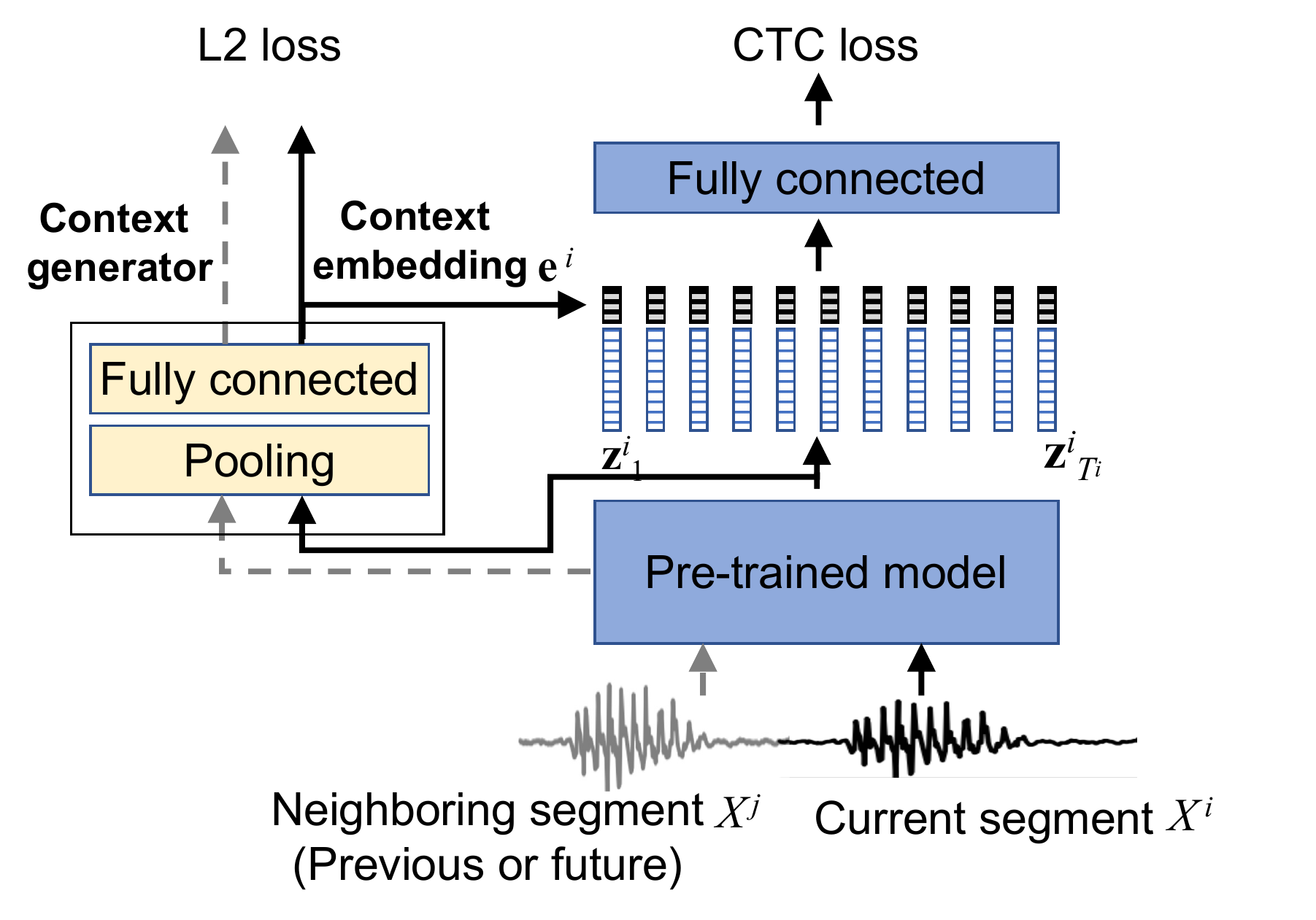}
         % \vspace{-0.5cm}
         \caption{Context-aware fine-tuning}
         \label{subfig:concept_method}
     \end{subfigure}
    \caption{Fine-tuning methods for self-supervised pre-trained speech models.}
    \label{fig:concept}
    \vspace{-0.5cm}    
\end{figure*}

\section{Context-aware fine-tuning}
\label{sec:Approaches}
\klremove{To recognize given speech segments, contextual information is a key factor to understand correctly.} \kledit{When recognizing a given speech segment, context information can be very helpful for correct recognition.}  Contextual information can be a word or phrase that is semantically related, or even higher-level \klremove{of} knowledge \kledit{(e.g., topic information)} that exists in the previous or \klremove{forthcoming}\kledit{following} speech. 
% To recognize given speech segments in general, the context of dialogue is a key factor to understand conversational speech more effectively and accurately. 
The self-attention layer of a transformer network is capable of capturing a long range of context, but the range is limited by the given speech segment. 
We hypothesize that an even longer range of global context can be captured in the neighboring speech segments beyond the current speech segment, whether they are in the past or the future.
In this section, we discuss our approach for leveraging previous or future speech segments while fine-tuning pre-trained speech models for any downstream task. 

Fig.~\ref{subfig:concept_baselineA} shows a typical example of fine-tuning a pre-trained speech model for ASR. The pre-trained speech model generates speech representations $\mathcal{Z}^i = \{\mathbf{z}^i_1,\mathbf{z}^i_2,...,\mathbf{z}^i_{T_i}\}$ from input audio $\mathcal{X}^i$ where $i$ is the current segment index of the speech stream and $T_i$ is the length in time \kledit{steps (frames) of the $i^\textrm{th}$} segment. The fully connected layer transforms the speech representations $\mathcal{Z}^i$ into \kledit{embeddings of dimensionality equal to target label vocabulary size, e.g.~the number of tokens, letters, or words.} \klremove{the target label dimensional embedding, e.g. size of token, letter, or word. }
The CTC loss can be used to fine-tune all of the model parameters. 
The fine-tuning dataset can be considerably smaller than would be needed for supervised ASR training from scratch.

\subsection{Context embedding using neighboring speech segments}
Given this fine-tuning framework, we introduce an approach to encode context information from \klremove{the}\kledit{a} neighboring speech segment.
The neighboring segment is a previous or future segment $\mathcal{X}^j$ where $j\ne i$.
Once the pre-trained model generates speech representations using the neighboring segment, a context module converts the variable-length speech representations into a context embedding vector using an attention-based pooling layer followed by a fully connected layer.
For example, when we use the nearest previous segment  $\mathcal{X}^{j}$, the speech representation $\mathcal{Z}^{j}$ is converted to a context embedding $\mathbf{e}^{j}$.
Lastly, this context embedding $\mathbf{e}^{j}$ is concatenated with the pre-trained model representations $\mathcal{Z}^i$ extracted from the current speech segment $\mathcal{X}^i$.
The new context-encoded speech representation is
\begin{equation}    
\mathcal{Z}^i_{j} = \{[\mathbf{z}^i_1,\mathbf{e}^{j}],[\mathbf{z}^i_2,\mathbf{e}^{j}],...,[\mathbf{z}^i_T,\mathbf{e}^{j}]\}.
\end{equation}
This framework can be fine-tuned for any downstream speech task as shown in Fig.~\ref{subfig:concept_baselineB} for the example of ASR fine-tuning.

\subsection{Context embedding using the current speech segment}

The context embedding provides a long-range global context that may help improve performance on downstream speech tasks. 
However, the context embedding needs the previous or future speech segments, and the model consumes twice the computational cost to generate speech representations. Additionally, the future speech segment is not available in a real-time speech system. 

To overcome these drawbacks, we propose a context generator module. This module consists of the same layers that extract the context embedding in Fig.~\ref{subfig:concept_baselineB}. The difference is that it generates the context embedding from the speech representations of the current speech segment. As shown in Fig.~\ref{subfig:concept_method}, the context generator module produces two different context embeddings, $\mathbf{e}^{j}$ using the neighboring speech segment and $\mathbf{e}^{i}$ from the current speech segment. The context loss $L_{context}$ is the L2 norm between the two \klremove{different} context embeddings:
\begin{equation}
    L_{context}=\| \mathbf{e}^{j}-\mathbf{e}^{i} \|_2.
\end{equation}
The context loss can be added to any loss for the downstream task with a weight $\alpha$. \klremove{This} \kledit{The} combined loss encourages the two embeddings to be closer in the latent embedding space.
Consequently, the proposed network consumes only the current speech segment during the forward pass while the neighboring speech segment is still needed during the training process. Thus, this model introduces nearly the same computational cost as the baseline model in Fig.~\ref{subfig:concept_baselineA}, as it only consumes the current speech segment during the forward pass. \kledit{Moreover, the speech system can be real-time, }\klremove{Moreover, there is no limitation for a real-time speech system }since it does not need the neighboring speech segment during inference.

\textbf{}\section{Experiments}
\label{sec:experiment}
\subsection{Experimental setup}
\subsubsection{Tasks and datasets}
We conduct experiments on ASR and SLU downstream tasks.
For the ASR task, we use SLUE-VoxCeleb~\cite{shon2022slue,nagrani2017voxceleb}, SLUE-VoxPopuli~\cite{shon2022slue,wang2021voxpopuli}, and LibriSpeech~\cite{panayotov2015librispeech}. 
The SLUE-VoxCeleb and SLUE-VoxPopuli datasets have 12.8 and 14.5 hours of fine-tuning data, respectively. We chose these datasets because they contain natural speech in the wild. Additionally, we also evaluate models on LibriSpeech using different amounts of labeled audio: 10m, 1h, 10h, and 100h using the splits provided in the Libri-light benchmark~\cite{kahn2020libri}.

For the SLU tasks, we use the SLUE-VoxPopuli dataset for NER and the SLUE-VoxCeleb dataset for SA.
NER involves detecting the named entities and their tags (types) in a given sentence. We evaluate an unordered list of named entity phrases and tag pairs predicted for each sentence using the F1 score.
SA refers to the task of classifying a given speech segment
as having negative, neutral, or positive sentiment. We evaluate SA using macro-averaged (unweighted) F1 scores.
For all evaluations, we use SLUE-Toolkit for the benchmark. For the SLUE datasets, we follow all pre-defined dataset splits and evaluation rules. 
% For the Libri-light benchmark, we used the same set defined in~\cite{kahn2020libri}.

\subsubsection{Fine-tuning}
The models are implemented using fairseq~\cite{ott2019fairseq} and SLUE-Toolkit. We \klremove{follow all}\kledit{use all of} the hyper-parameters defined in SLUE-Toolkit for SLUE dataset fine-tuning and fairseq for LibriSpeech fine-tuning except for additional hyper-parameters added in the proposed approach. We use the pre-trained wav2vec2.0 base model for all experiments. We use the same models in Fig.~\ref{fig:concept} for ASR and NER tasks with character targets plus word boundary tokens. The NER task has an additional 18 tokens to tag the start and end of each named entity. For the sentiment analysis task, we replace CTC loss with cross entropy loss, and the fully connected layer output dimension is changed to 3 to produce sentiment class output. For LM decoding, we follow the same setup as in~\cite{shon2022slue}.

\begin{figure}[t]
    \includegraphics[width=\linewidth]{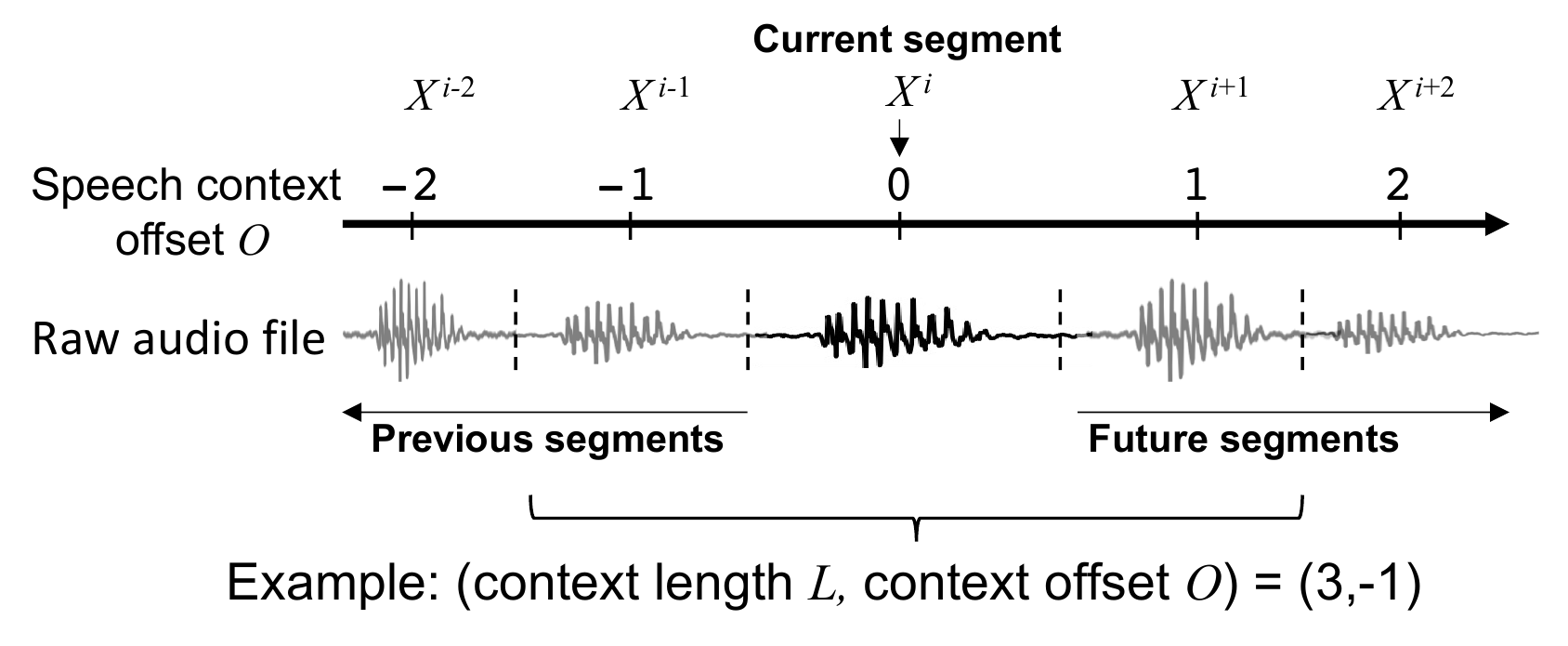}
    \vspace{-0.7cm}
    \caption{Example of \klremove{hyper-parameter:context}\kledit{the hyper-paremeters context} window length $\mathit{L}$ and context offset $\mathit{O}$.}
    \label{fig:example}
    \vspace{-0.5cm}
\end{figure}

\subsubsection{Hyper-parameters}
For the proposed approach, there are 4 new hyper-parameters: context offset $\mathit{O}$, context window length $\mathit{L}$, context loss weight $\alpha$, and context embedding size $\mathit{D}$. 

As shown in Fig.~\ref{fig:example}, the \textbf{context window length} $\mathit{L}$ and \textbf{context offset} $\mathit{O}$ determine how many previous/future context segments will be used. Each speech segment is processed in the pre-trained model individually. If there are multiple context speech segments, each segment's speech representation is generated individually and then concatenated. We explore 5 pairs of ($\mathit{L}$, $\mathit{O}$): $(2,0), (2,-1), (3,-2), (3,-1), (3,0)$.

\textbf{Context loss weight} $\alpha$ determines how much the context loss contributes to the total loss to be optimized. Considering the scale of the CTC loss, we experiment with $\alpha \in \{ 10^{-5}, 10^{-4}, \dots, 10^5\}$.

\textbf{Context embedding size} $\mathit{D}$ is the size of the context embedding that will be concatenated to the current speech representation as in Fig~\ref{fig:concept} (b) and (c).  The final speech representation has dimension $\mathit{D}$ plus the original pre-trained model representation dimension.  For example, $D+768$ is the final speech representation size when using the wav2vec2.0 base model.

In the proposed approach as in Fig~\ref{subfig:concept_method}, the context window length $\mathit{L}$ and context offset $\mathit{O}$ only affect the computational complexity of the training phase, while the context embedding size $\mathit{D}$ affects \kledit{both} the training and inference phases.\klremove{ as well.}

In Table~\ref{tab:lossweight},~\ref{tab:contextdim} and~\ref{tab:contextwindow}, we evaluate ASR performance using different hyper-parameter settings to find the optimal combination. All evaluation is done using the SLUE-VoxCeleb dev set as the evaluation set and the fine-tune set as a fine-tuning set. All fine-tuning is done 3 times with different random seeds, and we report the average\klremove{d} performance. We find that the optimal context loss weight $\alpha$ is between 0.1 and 10. For the context embedding size $\mathit{D}$, we observe no \klremove{more} performance improvement \kledit{when} using more than 32 dimensions. For the context window length $\mathit{L}$ and offset $\mathit{O}$, we observe slightly better performance when we include future context, i.e when $\mathit{L} + \mathit{O} > 1$. For training efficiency, we include 1 future context segment, i.e.~$\mathit{L}=2, \mathit{L}=0$. Consequently, we set the hyper-parameters ($\alpha$, $\mathit{D}$, $\mathit{L}$, $\mathit{O}$) to (10, 32, 2, 0) in the rest of the experiments. Using this setting, the overall parameter size is increased by 0.028\% (from 94.40M to 94.42M) compared to the baseline.
% 94,424,323 - 94,397,858 = 26,465
% 26465/94,397,858 = 0.028%

\begin{table}[!htp]\centering
\caption{Word error rate (WER) vs.~context loss weight $\alpha$ on the SLUE-VoxCeleb dev set. The context dimension size $\mathit{D}$, context window length $\mathit{L}$, and context window offset $\mathit{O}$ are set to 32, 2 and -1, respectively.}
\label{tab:lossweight}
\small
\begin{tabular}{lrrr}\toprule
\multirow{2}{*}{Weight $\alpha$} &\multicolumn{2}{c}{WER (on dev set)} \\\cmidrule{2-3}
&without LM &with LM \\\midrule
0.0001 &16.81 &13.01 \\
0.001 &16.82 &13.00 \\
0.01 &16.77 &12.97 \\
0.1 &16.73 &12.95 \\
1 &16.78 &12.97 \\
10 &\textbf{16.72} &\textbf{12.93 }\\
100 &16.82 &13.03 \\
1000 &17.14 &13.20 \\
10000 &18.37 &13.87 \\
100000 &76.58 &64.77 \\
\bottomrule
\vspace{-1cm}
\end{tabular}
\end{table}

\begin{table}[!htp]\centering
\caption{WER vs.~context dimension size $\mathit{D}$ on the SLUE-VoxCeleb dev set. The context loss weight $\alpha$, context window length $\mathit{L}$ and context window offset $\mathit{O}$ are set to 10, 2 and -1, respectively.}\label{tab:contextdim}
\small
%\begin{tabular}{lrrr}\toprule
\begin{tabular}{rrrr}\toprule
\multirow{2}{*}{Dimension $\mathit{D}$} &\multicolumn{2}{c}{WER (on dev set)} \\\cmidrule{2-3}
&without LM &with LM \\\midrule
2 &16.84 &13.00 \\
4 &16.88 &13.01 \\
8 &16.80 &12.94 \\
16 &16.81 &12.92 \\
32 &\textbf{16.66} &\textbf{12.88} \\
64 &16.75 &12.91 \\
128 &16.74 &12.99 \\
256 &16.73 &12.96 \\
\bottomrule
\vspace{-1cm}
\end{tabular}
\end{table}

\begin{table}[!htp]\centering
\caption{WER vs.~context window length $\mathit{L}$ and context window offset $\mathit{O}$ on the SLUE-VoxCeleb dev set. The context dimension size $\mathit{D}$ and context loss weight $\alpha$ are set to 32 and 10, respectively. (a): Fine-tuning baseline, (b): Context injection baseline, (c): Context-aware fine-tuning}\label{tab:contextwindow}
\small
%\begin{tabular}{lrrrrr}\toprule
\begin{tabular}{cccrrr}\toprule
\multirow{2}{*}{\makecell{Fine-tuning\\model}} &\multirow{2}{*}{\makecell{Window\\length $\mathit{L}$}} &\multirow{2}{*}{\makecell{Window\\offset $\mathit{O}$}} &\multicolumn{2}{c}{WER (on dev set)} \\\cmidrule{4-5}
& & &without LM &with LM \\\midrule
(a) &- &- &17.50 &13.30 \\\midrule
(b) &2 &-1 &16.76 &12.83 \\
(b) &2 &0 &16.77 &12.88 \\
(b) &3 &-2 &16.82 &12.86 \\
(b) &3 &-1 &\textbf{16.71} &12.88 \\
(b) &3 &0 &16.77 &\textbf{12.81} \\\midrule
(c) &2 &-1 &16.71 &12.92 \\
(c) &2 &0 &\textbf{16.67} &12.86 \\
(c) &3 &-2 &16.83 &12.90 \\
(c) &3 &-1 &16.77 &\textbf{12.84} \\
(c) &3 &0 &16.75 &12.87 \\
\bottomrule

\end{tabular}
\end{table}

\subsection{Results}

Tables~\ref{tab:sluescore_dev} and ~\ref{tab:sluescore} present the SLUE benchmark evaluation results on the dev and test sets. We choose the best system based on the performance on the dev set for each task and submit it to the benchmark website~\footnote{https://asappresearch.github.io/slue-toolkit}. For ASR, we observe 5\% and 8\% relative WER improvement on the test set without LM decoding for SLUE-VoxCeleb and SLUE-VoxPopuli, respectively. Interestingly, the proposed context-aware fine-tuning (c) performs competitively to the context injection baseline (b), which shows that we can obtain performance improvement without real context input during inference. With an LM, the proposed approach still consistently outperforms the baseline; however, the relative WER improvement is limited to 1.5\%.

\begin{table}[!tp]\centering
\caption{SLUE benchmark scores on \kledit{the} dev set. *:We found the SA baseline in~\cite{shon2022slue} is fairly low performing compared to our reproduced baseline model, so we report our reproduced number. VC:SLUE-VoxCeleb, VP: SLUE-VoxPopuli}\label{tab:sluescore_dev}
% \scriptsize
% \begin{tabular}{lrrrrrr}\toprule
\begin{tabular}{lcccccc}\toprule
\multirow{2}{*}{Model} &\multirow{2}{*}{\makecell{SLUE\\score}} &\multicolumn{2}{c}{ASR($\downarrow$)} &NER($\uparrow$) &SA($\uparrow$) \\\cmidrule{3-6}
& &VC &VP &VP &VC \\\midrule
(a) &60.3 &17.5 &17.5 &55.0 &43.3 \\
(a)* &62.2 &17.5 &17.5 &55.0 &49.2 \\
(b) &64.4 &16.7 &16.5 &60.0 &49.8 \\
(c) &64.4 &16.7 &16.4 &60.1 &49.7 \\\midrule
(a)+LM &65.9 &13.3 &12.0 &68.1 &43.3 \\
(a)*+LM &68.2 &13.3 &12.0 &68.1 &49.2 \\
(b)+LM &69.1 &12.7 &11.9 &69.8 &49.8 \\
(c)+LM &69.2 &12.8 &11.8 &70.1 &49.7 \\
\bottomrule
\vspace{-0.7cm}
\end{tabular}
\end{table}

\begin{table}[!tp]\centering
\caption{SLUE benchmark scores on \kledit{the} test set. *: our reproduced baseline.}\label{tab:sluescore}
% \scriptsize
% \begin{tabular}{lrrrrrr}\toprule
\begin{tabular}{lcccccc}\toprule
\multirow{2}{*}{Model} &\multirow{2}{*}{\makecell{SLUE\\score}} &\multicolumn{2}{c}{ASR($\downarrow$)} &NER($\uparrow$) &SA($\uparrow$) \\\cmidrule{3-6}
& &VC &VP &VP &VC \\\midrule
(a) &59.5 &20.9 &18.4 &49.6 &48.6 \\
(a)* &60.6 &20.9 &18.4 &49.6 &51.8 \\
(b) &62.4 &20.1 &17.2 &53.8 &52.1 \\
(c) &63.1 &20.0 &17.0 &55.0 &52.9 \\\midrule
(a)+LM &65.9 &16.1 &12.3 &63.4 &48.6 \\
(a)*+LM &67.0 &16.1 &12.3 &63.4 &51.8 \\
(b)+LM &67.4 &15.8 &12.1 &64.0 &52.1 \\
(c)+LM &67.7 &15.8 &12.1 &64.1 &52.9 \\
\bottomrule
\end{tabular}
\end{table}

The NER column shows the standard F1 score for the SLUE NER task following~\cite{shon2022slue}. The proposed approach shows a sizable 10\% relative improvement without an LM in the F1 score.
The SA column shows the SLUE sentiment task results in terms of F1 score. Although the relative F1 score improvement of the proposed approach is 2\% on the test set, it is still a noticeable improvement considering that in the previous study~\cite{shon2022slue}, the wav2vec2.0 large model gives only 3\% relative improvement compared to the base model.
The SLUE-score column shows the SLUE benchmark score results as an overall rating. We observe that the proposed model shows performance improvements with and without an LM, although the gain is even more significant \klremove{when} without an LM. 
Additionally, we also evaluate performance on the Libri-light ASR benchmark on the 10m, 1h, 10h, and 100h fine-tuning setups, and we observe a similar relative improvement as shown in Table~\ref{tab:asr_libri}.

\begin{table}[!htp]\centering
\caption{LibriSpeech dev/test set evaluation result \kledit{when} fine-tuning on Libri-light low-resource labeled data.}\label{tab:asr_libri}
% \scriptsize
\begin{tabular}{lcccc}\toprule
Model &dev-clean &dev-other &test-clean &test-other \\\midrule
\multicolumn{2}{l}{\textbf{10min labeled}} & & & \\
(a)\cite{baevski2020wav2vec} &46.1 &51.5 &46.9 &50.9 \\
(b) &41.7 &48.5 &42.5 &48.6 \\
(c) &41.7 &48.7 &42.6 &49.0 \\\midrule
\multicolumn{2}{l}{\textbf{1h labeled}} & & & \\
(a)\cite{baevski2020wav2vec} &24.1 &29.6 &24.5 &29.7 \\
(b) &21.9 &28.8 &22.0 &29.4 \\
(c) &21.8 &28.8 &22.0 &29.4 \\\midrule
\multicolumn{2}{l}{\textbf{10h labeled}} & & & \\
(a)\cite{baevski2020wav2vec} &10.9 &17.4 &11.1 &17.6 \\
(b) &9.2 &16.5 &9.1 &16.9 \\
(c) &9.0 &16.4 &9.1 &16.7 \\\midrule
\multicolumn{2}{l}{\textbf{100h labeled}} & & & \\
(a)\cite{baevski2020wav2vec} &6.1 &13.5 &6.1 &13.3 \\
(b) &5.4 &13.3 &5.2 &13.0 \\
(c) &5.1 &13.3 &5.3 &13.0 \\
\bottomrule
\end{tabular}
\end{table}

% For librispeech, we fine-tuned the model 3 times with random seed and reported averaged WER without LM decoding. We observed relative WER improvement on 8\%~13\% using the proposed approach on the test-clean. In the test-other set, we observed 

\begin{figure}[t]
\centering
    \includegraphics[width=0.7\linewidth]{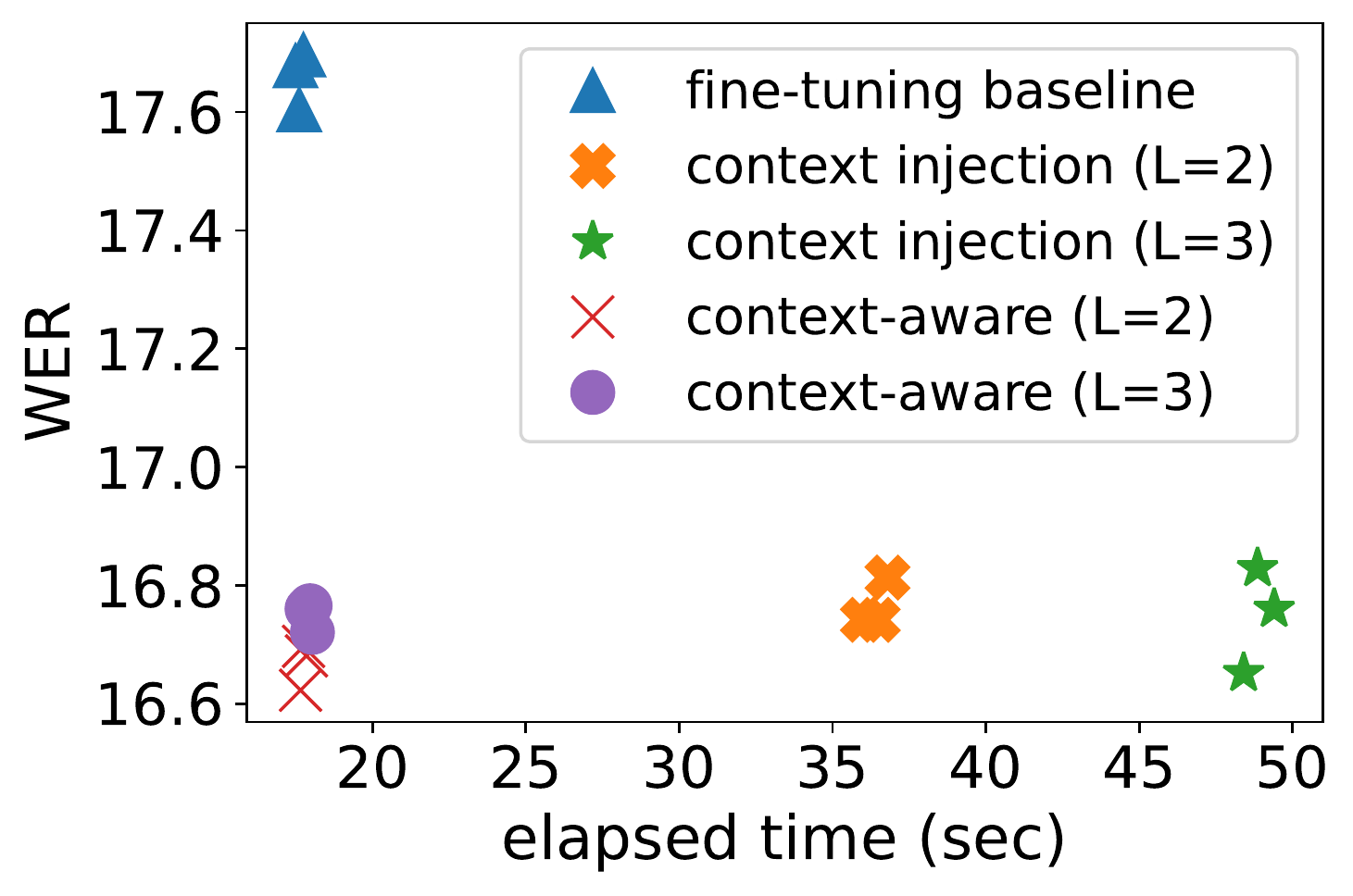}
    \caption{WER versus inference time comparison. Note that audio loading stage was not counted in this table and offset $\mathit{O}$ is set to 0.}
    \label{fig:speed}
    \vspace{-10pt}
\end{figure}

\subsection{Discussion}
The experimental results show that contextual information from neighboring segments is helpful for extracting information from the current segment. The hyper-parameter study shows that using the future neighboring segment is slightly better than using the past neighboring segment as shown in Table~\ref{tab:contextwindow}. Using future neighboring segments is limited in real-time or streaming setups, while context-aware fine-tuning is free from this limitation. 
Moreover, the proposed context-aware fine-tuning approach does not sacrifice inference speed as shown in Figure~\ref{fig:speed} and performs \kledit{similarly} to the context injection baseline.
% In all experiments, the performance gain with LM decoding is smaller than the gain without LM. 
% However, using LM shallow fusion includes additional model parameters and complexity in the software interface. 
% In addition, LM shallow fusion cannot consider future context while context-aware fine-tuning can utilize future segments as well that might give exclusive context information.
% A likely reason is that contextual information from neighboring segments has some redundancy with the LM. Also, the LM-related hyper-parameters could be sub-optimal, but we did not explore that thoroughly in this study.
%One limitation of this study is that we have only applied the method in fine-tuning. Since the context loss can be used with self-supervision, it can naturally be applied in the pre-training stage. We expect that matching the receptive field in the pre-training and fine-tuning stages could improve performance on downstream tasks by learning contextual information.  

% \vspace{-0.1cm}
% \vspace{-4mm}
\section{Conclusions}
\label{sec:conclusions}
Our experiments show that the proposed context-aware fine-tuning approach, which utilizes the contextual information from surrounding audio segments during fine-tuning, can improve performance with only a tiny overhead to inference. One limitation of this study is that we have only applied the method in fine-tuning. Since our proposed context loss can be used \klremove{under self-supervision,}\kledit{in a self-supervised setting,} it can naturally be applied in the pre-training stage. Matching the "receptive field" by applying the proposed method in the pre-training and fine-tuning stages may further improve performance on downstream tasks. Additional future work includes extending context-aware fine-tuning to other SLU tasks such as speech summarization~\cite{sharma2021speech} and intent classification~\cite{bastianelli2020slurp,lugosch2019speech}.

\clearpage

\clearpage
\vfill\pagebreak

\bibliographystyle{IEEEbib}
{\footnotesize \bibliography{refs}}
% \bibliography{refs}

% \ifdraft
% \ifarxiv
% \onecolumn 
% \normalsize
% % \input{sections/appendix}
% \fi

\end{document}